\documentclass[11pt]{article}
\usepackage{epsfig}
\newcommand{\sss}{\vspace{.2in}}
\newcommand{\sms}{\vspace{.1in}}
\def\br{\begin{eqnarray}}
\def\er{\end{eqnarray}}
\def\be{\begin{equation}}
\def\ee{\end{equation}}
\def\lb{\lbrack}
\def\rb{\rbrack}

\def\>{\rangle}              %%  > for `ket'
\def\<{\langle}              %%  < for `bra'
\def\({\left(}
\def\){\right)}
\def\[{\left[}
\def\]{\right]}
\def\v{\vert}                     %% vertical bars
\def\ra{\rightarrow}
\def\lra{\longrightarrow}

\def\wti{\widetilde}
\def\eq{\!\!\!\! &=& \!\!\!\! }
\def\jp{J_{+}}
\def\jm{J_{-}}
\def\j3{J_3}
\def\Ad{{\cal A}^{\dagger}}
\def\A{{\cal A}}

\def\half{\frac{1}{2}}
\def\id{i\partial_\phi}

\newcommand{\nn}{\nonumber}
\newcommand{\cosec}{{\rm \,cosec}}
\newcommand{\sech}{{\rm \,sech}}
\newcommand{\cosech}{{\rm \,cosech}}
\begin{document}
~\hfill{\footnotesize UICHEP-TH/00-2,~~\today}
\sss
\begin{center}
{\Large {\Large \bf New Solvable Singular Potentials}}
\end{center}
\vspace{.5in}
\begin{center}
{\large{\bf
   \mbox{R. Dutt}$^{a,}$\footnote{rdutt@cal2.vsnl.net.in},
   \mbox{A. Gangopadhyaya}$^{b,}$\footnote{agangop@luc.edu, asim@uic.edu},
   \mbox{C. Rasinariu}$^{c,}$\footnote{crasinariu@popmail.colum.edu, costel@uic.edu} and
   \mbox{U. Sukhatme}$^{d,}$\footnote{sukhatme@uic.edu}
 }}
\end{center}
\vspace{.6in}
{\small
\noindent
a) \hspace*{.11in}
Department of Physics, Visva Bharati University, Santiniketan 731235, India\\
b) \hspace*{.11in}
Department of Physics, Loyola University Chicago, Chicago, Illinois 60626 \\
c) \hspace*{.11in}
Department of Science and Mathematics, Columbia College Chicago, Chicago, Illinois 60605 \\
d) \hspace*{.11in}
Department of Physics (m/c 273), University of Illinois at Chicago, Chicago, Illinois 60607 \\
}
\begin{abstract}
We obtain three new solvable, real, shape invariant potentials starting from the
harmonic oscillator, P\"oschl-Teller I and P\"oschl-Teller II potentials on the
half-axis and extending their domain to the full line, while taking special care to
regularize the inverse square singularity at the origin. The regularization
procedure gives rise to a delta-function behavior at the origin. Our new systems
possess underlying non-linear potential algebras, which can also be used to
determine their spectra analytically.
\end{abstract}

\newpage
\noindent{\bf I. Introduction:}\sms

In recent years, several authors have investigated the eigenstates of complex
potentials \cite{Bender}, especially those with PT symmetry and real spectra. In
particular, potentials obtained by replacing the real coordinate $x$ by a complex
variable $x+ic$ in a class of exactly solvable shape invariant potentials has been
considered by Znojil \cite{Znojil}. For these problems, which are defined on the
whole real line,  an extension to the complex domain was made in order to avoid an
inverse-square singularity at the origin. The price one pays for taming the
singularity is to deal with a complex potential. However, it was argued that owing
to the PT-symmetric nature of the potential, the eigenvalues were still real.  As an
explicit example, it was shown \cite{Znojil} that a new exactly solvable complex
harmonic-oscillator like potential with two shifted sets of equally spaced energy
levels could be generated. The same technique was also applied to explicitly  get
the eigenstates of complex P\"oschl-Teller I and P\"oschl-Teller II like
potentials\cite{Znojil2}.

One of the purposes of this paper is to show that potentials with an inverse square
singularity at the origin do not necessarily call for moving into the complex
domain. In fact, we can obtain the spectra of refs. \cite{Znojil,Znojil2} simply by
judicious application of the formalism of supersymmetric quantum mechanics.
Specifically, the spectrum described in ref. \cite{Znojil} for the
harmonic-oscillator like potential is identical to that previously found by us
\cite{equishift}, where the discussion focused on real potentials with two sets of
equally spaced eigenvalues. In this paper, we extend previous results and also find
new real but singular potentials corresponding to the P\"oschl-Teller I and
P\"oschl-Teller II potentials. Our potentials are shape invariant \cite{Infeld}, and
consequently their exact spectra can be obtained by standard algebraic procedures
followed in supersymmetric quantum mechanics. We also establish that these singular
potentials possess an interesting underlying non-linear potential
algebra \cite{Wu,Alhassid,Gangopadhyaya_proc,ASIM,Balantekin}.  Explicit
representations of the generators are given, which provide an alternative algebraic
approach to determine the spectrum.

For completeness, we provide in Sec. 2 a brief review of supersymmetric quantum
mechanics (SUSYQM)\cite{Witten,Cooper}. In Sec. 3, we present our framework for
generating new shape invariant potentials starting from well known solvable problems
with an inverse square singularity at the origin. We show that if the coefficient of
this  singularity is restricted within a narrow range, one can enlarge the domain of
the potential to the negative real axis while maintaining unbroken supersymmetry and
shape invariance. Working with an explicit example of a harmonic oscillator with an
inverse square singularity and using the formalism of supersymmetric quantum
mechanics, we show that such an extension necessitates the introduction of a
$\delta$-function at the origin \cite{equishift} which eventually yields a
non-equidistant spectrum for the system. We show that similar extensions can be made
for P\"{o}schl-Teller I and P\"{o}schl-Teller II potentials as well and, thus
generate new shape invariant potentials.
%---------
We explicitly derive their eigenenergies and eigenfunctions. For these potentials,
it is important to note that the eigenenergies depend on two parameters, {\it both}
of which get transformed in the shape invariance condition, in contrast to previous
work on shape invariance.
%---------
In Sec. 4, we study the potential algebra underlying these systems and generate
their spectrum by algebraic means.  \sss

\noindent {\bf 2. Supersymmetric Quantum Mechanics:}\sms

In supersymmetric quantum mechanics \cite{Cooper}, taking $\hbar=2m=1$, the partner
potentials $V_{\pm}(x,a_0)$ are related to the superpotential $W(x,a_0)$ by
\begin{equation} \label{vpm}
V_{\pm}(x,a_0)=W^2(x,a_0) \pm W'(x,a_0)~~,
\end{equation}
where $a_0$ is a set of parameters. It is assumed that the superpotential $W(x)$ is
continuous and differentiable. The corresponding Hamiltonians $H_{\pm}$ have a
factorized form
\be \label{hpm}
H_-=\Ad \A~,~H_+=\A \Ad~,~\A = {d \over {dx}} +W(x)~,~ \Ad=-{d \over {dx}} +W(x).
\ee
We consider the case of unbroken supersymmetry and take ${\psi_0} \sim \exp\left( -
\int^x W(y) dy \right)$ to be normalizable. This is clearly the nodeless zero energy
ground state wave function for $H_-$, since $\A \psi_0 = 0$.

The Hamiltonians $H_+$ and $H_-$ have exactly the same eigenvalues except that $H_-$
has an additional zero energy eigenstate. More specifically, the eigenstates of
$H_+$ and $H_-$ are related by
\be
\label{susy}
E^{(-)}_0 = 0~,~~~E^{(+)}_{n-1}\;=\;E^{(-)}_{n},~~~ \psi^{(+)}_{n-1} \propto \A \,\psi^{(-)}_{n}~~,~~~\Ad \,\psi^{(+)}_{n} \propto \psi^{(-)}_{n+1}  ~~, \quad   n=1,2, \ldots.
\ee

Supersymmetric partner potentials are called shape invariant if they both have the
same $x$-dependence upto a change of parameters $a_1=f(a_0)$ and an additive
constant which we denote by $R(a_0)$\cite{Dutt86,dutt}. Often, it is convenient to
write this constant in the form of $g(a_1)- g(a_0)$. The shape invariance condition
is
\begin{equation}
\label{sipv}
V_+(x,a_0) = V_-(x,a_1) + R(a_0)=V_-(x,a_1) + g(a_1)- g(a_0)~.
\end{equation}
The property of shape invariance permits an immediate analytic determination of energy eigenvalues \cite{Infeld,dutt}, and eigenfunctions \cite{Dutt86}. If the change of parameters
$a_0 \ra a_1$ does not break supersymmetry, $H_-(x,a_1)$ also has a zero energy ground state and the corresponding eigenfunction is given by $\psi_0^{(-)}(x, a_1) \propto \exp \(- \int^x_{x_0} W(y,a_1) dy \)$. Now using eqs. (\ref{susy},\ref{sipv}) we have
\be
E_1^{(-)} = R(a_0)~,~~ \psi_1^{(-)}(x,a_0) = \Ad(x,a_0)\, \psi_0^{(+)}(x,a_0)=
\Ad(x,a_0) \,\psi_0^{(-)}(x,a_1)~.
\ee
Thus for an unbroken supersymmetry, the eigenstates of the potential $V_-(x)$ are:
\br \label{eq4}
E_0^{(-)} \eq 0~,~E_n^{(-)}=\sum_{k=0}^{n-1}R(a_k)=\sum_{k=0}^{n-1} \[g(a_{k+1})- g(a_{k})\]=g(a_n)- g(a_0)~,\\
\psi_0^{(-)} &\!\!\!\!\propto&\!\!\!\! e^{- \int^x_{x_0} W(y,a_0) dy}~,~
\psi_n^{(-)}(x,a_0)=\left[-\frac{d}{dx}+W(x,a_0)\right]
\psi_{n-1}^{(-)}(x,a_1)~,~~(n=1,2,3,\ldots)~.\nonumber
\er
These formulas are valid provided the change of parameters $a_1 = f(a_0)$ maintains
unbroken supersymmetry. In previous work on shape invariant potentials, changes of
parameters corresponding to  translation $a_1 = a_0 + \beta$ \cite{dutt} and scaling
$a_1 = q a_0$ with $0 <q \le 1$ \cite{Barclay} have been discussed. However, a
reflection change of parameters $a_1 = -a_0$, even if it maintained shape
invariance, was not acceptable since it could not maintain unbroken supersymmetry
for the hierarchy of potentials built on $H_-$.\sss

\noindent{\bf 3. New Singular Shape Invariant Potentials:}\sms

The methodology of this paper for obtaining new shape invariant potentials is as
follows. One begins with a known shape invariant potential, defined for $x \ge 0$,
which has an inverse square singularity $\lambda/x^2$ at the origin. This potential
is fully solvable, with eigenfunctions which vanish at the origin. One now considers
extending the domain to also include the region $x<0$. This extension is possible
only if $-1/4 < \lambda < 3/4$.  If the strength of the singular term is restricted
to be in this limited domain, the singularity is called ``soft", and the potential
is said to be ``transitional"\cite{Gangopadhyaya_inv_sq}.  We shall show explicitly
how a new change of parameters corresponding to the reflection $a_1 = -a_0$ is now
admissible, since it maintains both shape invariance and unbroken supersymmetry,
while still keeping the partner potentials in the soft singularity domain. We can
then obtain eigenspectra using the shape invariance formalism. As explicit examples,
we present detailed analyses for the harmonic oscillator, P\"oschl-Teller I and
P\"oschl-Teller II potentials.\sss

\noindent{\bf
(a) New shape invariant potential obtained from the harmonic oscillator potential.}

Consider a particle constrained to move in a three dimensional harmonic oscillator
potential
\begin{equation} \label{vr}
V_-(x,l,\omega)=   \frac{1}{4} ~\omega^2 x^2 +\frac{l(l+1)}{x^2} +\left(l-\frac{1}{2} \right)\omega  ~~,~~(0 < x <\infty)~.
\end{equation}
This potential is generated from the superpotential
\begin{equation}
\label{wx}
W(x,l,\omega)=   \frac{1}{2} ~\omega x +\frac{l}{x}~~;\quad l<0 ~.
\end{equation}
The supersymmetric partner potential is
\be
\label{vp}
V_+(x,l,\omega)= \frac{1}{4} ~\omega^2 x^2 +\frac{l(l-1)}{x^2} +  \left(l+\frac{1}{2} \right)\omega~~.
\ee
These two partner potentials are shape invariant since $V_+(x,l,\omega)$ can be
written as $V_-(x,l-1,\omega) + R(l,\omega)$. Here, the remainder
$R(l,\omega)=2\omega$ is independent of the parameter $l$. This yields an
equidistant spectrum $E_n=2n\omega$ for the harmonic oscillator. However, there is
another change of parameters that also maintains shape invariance between these two
partner potentials, namely
$$V_+(x,l,\omega)= V_-(x,-l,\omega) + R'(l,\omega)~~,$$
$R'(l,\omega) = (2l+1)\omega$.

However, it is important to point out that since we are at present constrained to be
on the half-axis $x>0$, this second change of parameters, $(l,\omega)
\longrightarrow (-l,\omega)$ is not acceptable for $l<0$. Neither of the two zero
energy solutions $\psi_0^{(\pm)}(x,-l,\omega) \propto \exp\( \pm \int^x
W(x,-l,\omega)\,dx  \)$  is normalizable and hence supersymmetry is spontaneously
broken. As shown in sec. 2, the solvability of shape invariant systems crucially
depends upon superpotentials retaining unbroken supersymmetry when parameters are
transformed, that is, it is essential that $V_-(x,a_1)$ be a potential with unbroken
supersymmetry.

Let us now consider the same superpotential with an extension of the domain to the
entire real axis. The asymptotic values of the superpotential are given by the
$\half \omega x$ term at $x \rightarrow \pm \infty $, which is independent of the
parameter $l$. Thus the asymptotic behavior of the ground state wave function is
dictated by the $\omega x$-term and is not affected by flipping of the value of $l$
in the $\frac{l}{x}$-term of the superpotential. Thus, in contrast with the
half-axis case, supersymmetry now remains unbroken even with the change of
parameters $(l,\omega) \longrightarrow (-l,\omega)$, and hence this transformation
is allowed to generate new shape invariant potentials with richer spectra. This
leads to
\be \label{energy}
E_n=n\omega + 2l\omega P_n~,~P_n \equiv [ 1-(-1)^n ]/2~.
\ee
This new shape invariance yields a new set of eigenenergies superimposed on the old
equidistant spectrum and are shown in Fig. 1.
\begin{figure}[ht] \label{fig1}
    \centering
    \epsfig{file=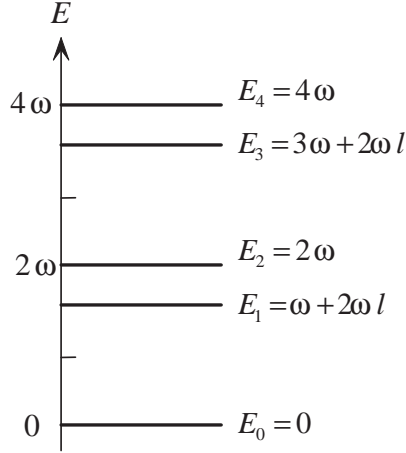,width=2.2in}
\caption{Energy eigenvalues corresponding to eq. (\ref{energy})}
\end{figure}

We now focus on the region near $x=0$. In SUSYQM, it is important that the
superpotential $W(x,a_0)$ be a continuous and differentiable function. In our
example, the above requirement is satisfied everywhere except at the point $x=0$,
where the superpotential of eq. (\ref{wx}) has an infinite discontinuity. Such a
discontinuity is not acceptable, and needs regularization. Consider a regularized,
continuous superpotential $\widetilde {W}(x,a_0,\epsilon)$ which reduces to
$W(x,a_0)$ in the limit $\epsilon \to 0$. One such choice is
\begin{equation} \label{Wt}
\widetilde {W}(x,a_0,\epsilon) = W(x,a_0)~f(x,\epsilon)
\end{equation}
where
\begin{equation}
f(x,\epsilon)= \tanh ^2 {\frac{x}{\epsilon}}~~.
\end{equation}
The moderating factor $f$ provides a smooth interpolation through the discontinuity,
since it is unity everywhere except in a small region of order $\epsilon$ around
$x=0$. In this region, $\widetilde {W}(x,a_0,\epsilon)$ is linear with a slope
$l/\epsilon^2$. The potential\footnote{At this point one may wonder whether we have
lost our cherished shape invariance due to the introduction of this moderating
factor. In Appendix A, we show that the shape invariance indeed remains intact in
the limit $\epsilon\to 0$, and so does the solvability of the model.} $\widetilde
{V}_-(x,a_0,\epsilon)$ corresponding to the superpotential $\widetilde
{W}(x,a_0,\epsilon)$ is
\begin{equation} \label{Vte}
\widetilde {V}_-(x,a_0,\epsilon) = \widetilde {W}^2(x,a_0,\epsilon)-
\widetilde{W}'(x,a_0,\epsilon)~.
\end{equation}
In the limit $\epsilon\to 0$, $\widetilde {V}_-(x,a_0,\epsilon)$ reduces to
\begin{equation} \label{Vt}
\widetilde{V}_-(x,a_0)=
V_-(x,a_0)-4~{W}(x,a_0)~\frac{x}{|x|}
~\delta(x)~,
\end{equation}
where we have used
${\lim}_{\epsilon\to 0}
~\frac{1}{2\epsilon} \sech^2\frac{x}{\epsilon} =\delta(x)$ and
${\lim}_{\epsilon\to 0}
~\tanh\frac{x}{\epsilon}= \frac{x}{|x|} $.
Thus we see that the potential $\widetilde {V}_-(x,a_0)$ has an
additional singularity at the origin over ${V}_-(x,a_0)$ given by
$\Omega(x) \equiv - 4 \,l \,\frac{~\delta(x)}{|x|}$.\\

\begin{figure}[ht] \label{fig2}
    \centering
    \epsfig{file=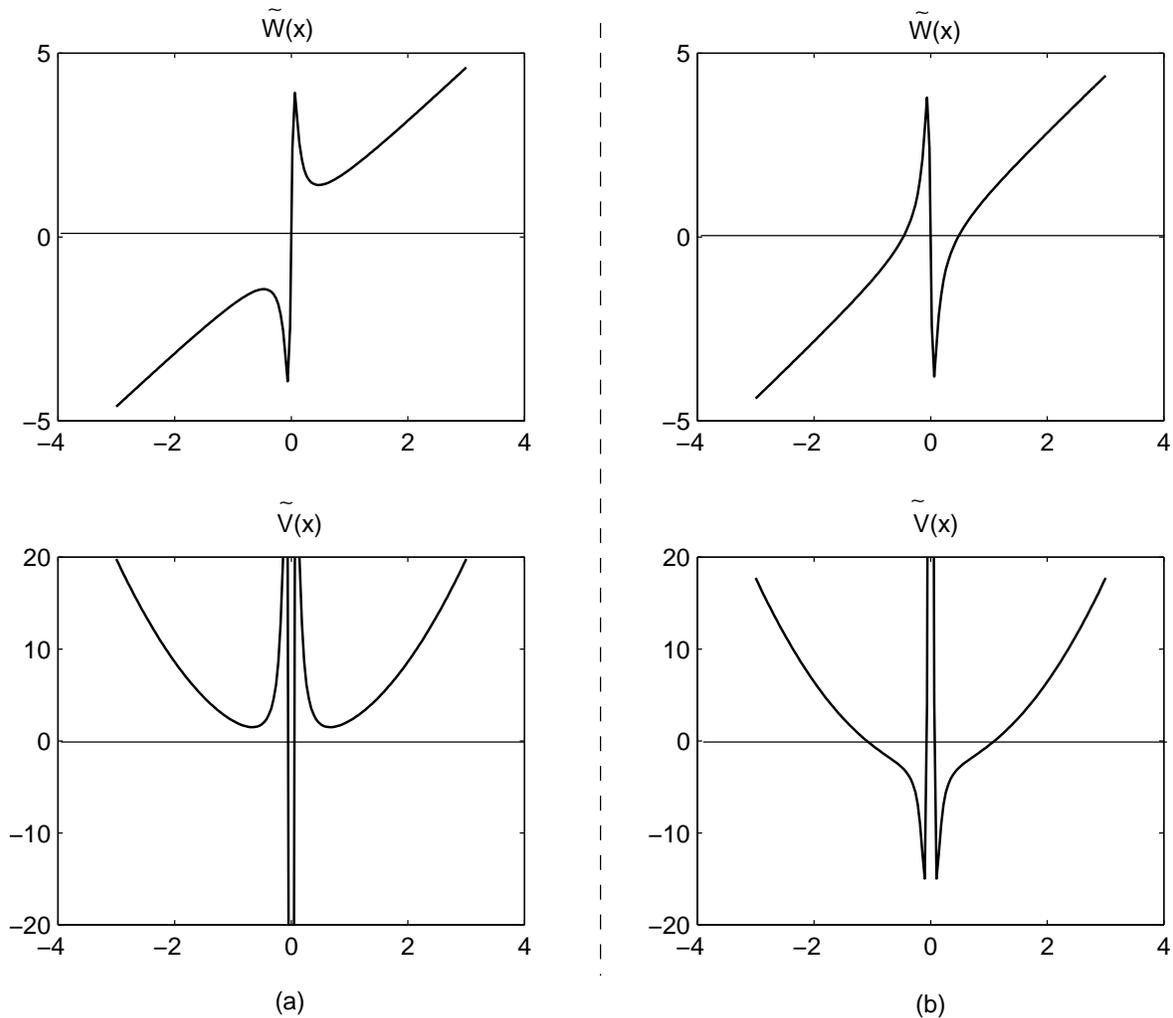,width=6.2in}
\caption{The superpotential $\widetilde W$ of eq. (\ref{Wt}) and the corresponding
potential ${\widetilde V_-}$ of eq. (\ref{Vte}) for the two cases
(a) $l \ge 0$ and (b) $l \le 0$.}
\end{figure}

Note that in the potential shown in Fig. 2(a), the $\delta$-function singularity is
instrumental in producing a bound state at $E_0=0$.

Naively, in the limit $\epsilon \to 0$, the potential of eq. (\ref{Vt}) appears
identical to a three dimensional oscillator with a frequency $\omega $ and angular
momentum $l$. However, there are some more subtle but important differences. First,
it is defined over the entire real axis ($-\infty<x<\infty$) and not just the half
line. For a proper communication between the two halves, we must have a ``softness"
of the inverse square term. Normalizability of the wave function requires that the
coefficient $\lambda$ of the inverse square term be in the transition region
$-{1\over 4} < \lambda < 3 \over 4$ \cite{Gangopadhyaya_inv_sq}. More specifically,
for $(l>0)$, one has $0<l(l+1)<\frac{3}{4}$ and for $(l<0)$ one has
$-\frac{1}{4}<l(l+1)<0$. The important special case of the one dimensional harmonic
oscillator has $l=0$: it corresponds to $l(l+1)=0$ and no $x^{-2}$ singularity. For
transition potentials, both solutions of the Schr\"odinger equation are square
integrable at the origin. Therefore, both are acceptable square integrable solutions
of the Schr\"odinger equation, and must be retained to form a complete set.
Eigenstates for the potential $\widetilde{V}_-(x,a_0)$  can be obtained from eq.
(\ref{eq4}). The lowest four are
\sss
\begin{eqnarray}
\label{eigenstates}
&E_0=0;~~~~&
\psi_0 ~\propto ~x^{-l} ~
e^{-{1\over 4} \omega x^2}~,\\
&E_1=(2l+1)\omega;~~~&
\psi_1 ~\propto ~x^{1+l} ~
e^{-{1\over 4} \omega x^2}~,\nonumber \\
&E_2= 2\omega~;~&
\psi_2 ~\propto ~
\left( 2l-1+ \omega x^2 \right)
x^{-l}
e^{-{1\over 4} \omega x^2}~,\nonumber \\
&E_3=2\omega+(2l+1)\omega~;&
\psi_3 ~\propto ~
\left( -2l-3 + \omega x^2 \right)
~x^{1+l} ~
e^{-{1\over 4} \omega x^2}~.\nonumber
\end{eqnarray}
General expressions for these eigenfunctions and corresponding eigenenergies are
\begin{eqnarray}
E_{2n}=2n \omega~,~ ~~~~~&& \psi_{2n} \propto
x^{-{l}} e^{-{1\over 4}\omega x^2} L_n^{-l-{1\over 2}}
\left[\frac{\omega  x^2}{2}\right]~, \nonumber \\
E_{2n+1}=2n\omega+ (2l+1)\omega~,~
&&
\psi_{2n+1} \propto
x^{1+{l}} e^{-{1\over 4}\omega x^2} L_n^{{l}+{1\over 2}}
\left[\frac{\omega x^2}{2}\right] ~~,
\label{spectrum-2}
\end{eqnarray}
\sss
\noindent where $L_n$ are the standard Laguerre polynomials.\sms

\sss

\noindent{\bf
(b) New shape invariant potential obtained from the P\"oschl-Teller I potential.}

As a second example, we consider the P\"oschl-Teller I superpotential
\be \label{wpt1}
W(x,A,B) = A \tan x-B \cot x~~;~~~~0<x<\pi/2~~.
\ee
The supersymmetric partner potentials are then given by
\be
V_-(x,A,B) = A(A-1) \sec^2x +B(B-1) \cosec^2x - (A+B)^2~~,
\ee
and
\be
V_+(x,A,B) = A(A+1) \sec^2x +B(B+1) \cosec^2x - (A+B)^2~~.
\ee
Here, $A$ and $B$ are both positive in order for $V_-(x,A,B)$ to have a zero energy
ground state.

Again, one can readily check that there are two possible relations between
parameters such that above two potentials exhibit shape invariance. One of them is
the conventional $(A \ra A+1,~~ B \ra B+1)$. The second possibility is $(A \ra
A+1,~~ B \ra -B)$. As explained in the previous section, this second relationship
breaks supersymmetry on $(0, \pi/2)$ domain and it is allowed only if the domain of
$x$ is extended to range $-\pi/2 <x<\pi/2$. The first transformation among
parameters $(A \ra A+1,~~ B \ra B+1)$ has been studied extensively in the
literature. It is the second transformation that yields new results and will be
considered here.  Thus, the relationship among parameters that we consider is,
$(A_{k+1}= A_k+1,~~ B_{k+1}= -B_k)$. This potential also requires a careful analysis
in the vicinity of $x=0$, where two half-axes are being sewed together. Again, the
need of continuity and differentiability of the superpotential requires its
regularization, as was done in eq. (\ref{Wt}) for the harmonic oscillator. A similar
analysis then leads to a new singular shape invariant potential
\be\label{20}
\wti{V}_-(x,A,B) = \[ A(A-1) \sec^2x +B(B-1) \cosec^2x
    - (A+B)^2\] +4~B \cot x~\frac{x}{|x|}
~\delta(x)~
.
\ee
This potential obeys the shape invariance condition:
\be
\wti{V}_+(x,A,B) = \wti{V}_-(x,A+1,-B) + (A+1-B)^2-(A+B)^2~~,
\ee
and its eigenvalues and eigenfunctions are given by
\br
&& E_0=0~, \nn \\
&& \psi_0 \propto \cos^Ax ~\sin^Bx~,
\nn \\
&& E_1=(A+1-B)^2-(A+B)^2~, \nn \\
&& \psi_1 \propto \cos^Ax ~\sin^{-B-1}x~
[(2B-1)\cos^2x+1]~,
\nn \\
&& E_2=(A+2+B)^2-(A+B)^2~,  \nn \\
&& \psi_2 \propto \cos^Ax ~\sin^Bx~
[(4B+2)\cos^4x-(6B+3)\cos^2x+1]~,
\nn \\
&& E_3=(A+3-B)^2-(A+B)^2~,  \nn \\
&& \psi_3 \propto \cos^Ax~ \sin^{-B-1} x~
\[(-8B^2+16B-6)\cos^6x+(12B^2-32B+13)\cos^4x+(20B-8)\cos^2x+1\]~.\nn
\er
Thus, the general formula for eigenvalues is
\be\label{PT-Ienergy}
E_n=(A+n+(-1)^n B)^2-(A+B)^2~~.
\ee
The eigenspectrum is shown in Fig. 3 and Fig. 4.
\begin{figure}[ht] \label{fig3}
    \centering
\epsfig{file=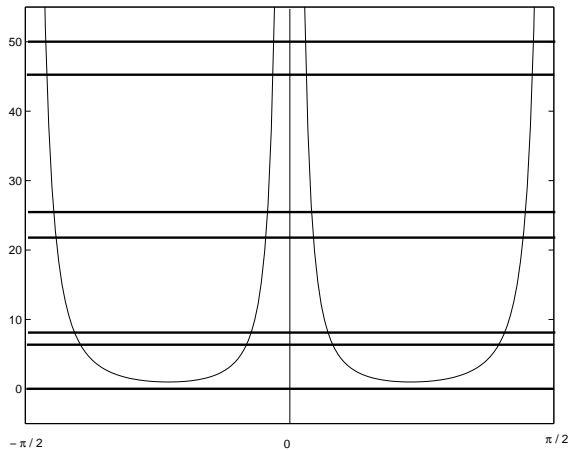,width=3.0in}
\caption{The potential of eq. (\ref{20}) for $A=1.5$
        and $B=-1/3$ and its energy spectrum.}
\end{figure}

\begin{figure}[ht] \label{fig4}
    \centering
    \epsfig{file=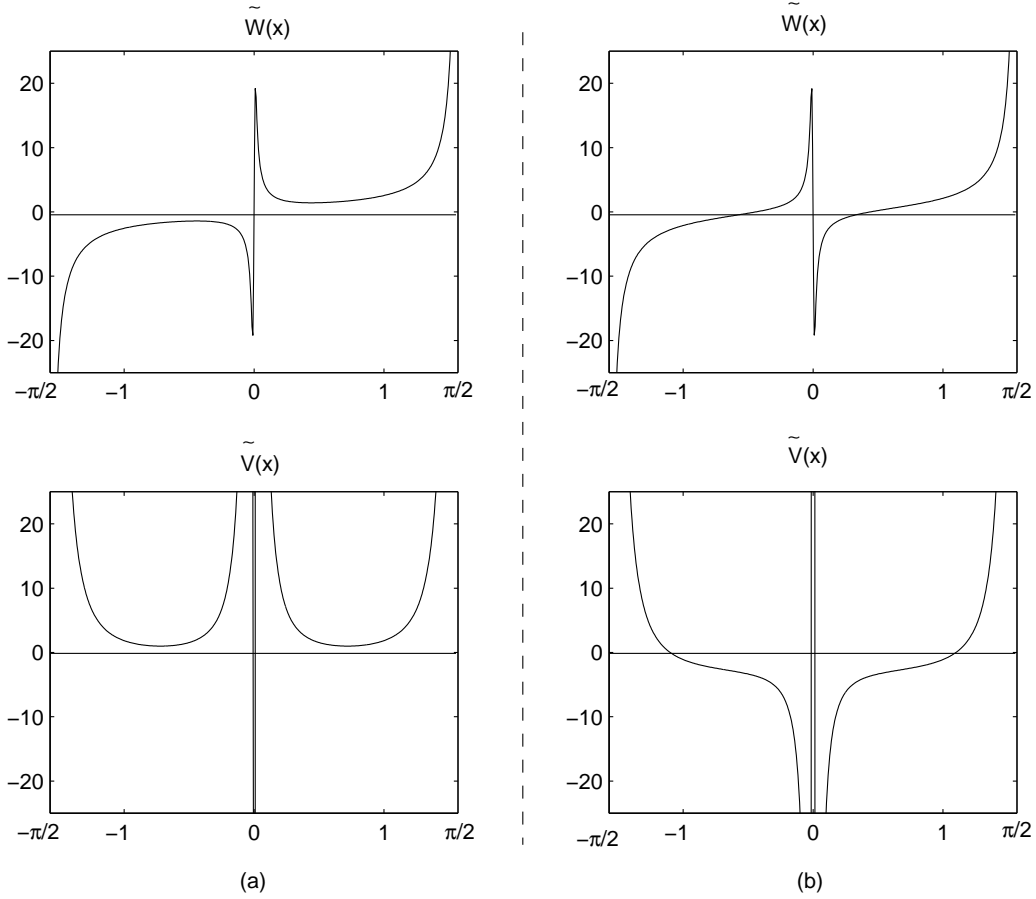,width=5.5in}
\caption{The superpotential $\widetilde W$ potential ${\widetilde V_-}$
from P\"oschl-Teller for $A=1.5$ and $B=-1/3$.}
\end{figure}
Note that, to avoid level crossing, we must have $E_n>E_{n-1}$. This leads to the
constraint $-\half < B <\half$. Interestingly, it is the same constraint that one
needs for the normalizability of the wavefunction at the origin and hence to the
possibility of communication between regions ($-\half\pi,0)$ and ($0,\half\pi)$ of
the domain.

\sss

\noindent{\bf (c) New shape invariant potential obtained from the P\"oschl-Teller
II potential.}

The last example that we consider is that of the P\"oschl-Teller II potential
described by
\be \label{wpt2}
W(x,A,B) = A \tanh x-B \coth x~~;~~~~0<x<\infty~~.
\ee
Here, $A$ and $B$ both need to be positive and satisfy the condition $A>B$ for the
potential $V_-(x,A,B)$ to have a zero energy ground state and to ensure unbroken
supersymmetry. The supersymmetric partner potentials are then given by
\be
V_+(x,A,B) = -A(A-1) \sech^2x +B(B+1) \cosec^2x + (A-B)^2
\ee
and
\be
V_-(x,A,B) = -A(A+1) \sec^2x +B(B-1) \cosec^2x + (A-B)^2~~.
\ee

Here too we have two possible relations  between parameters for these potentials to
be shape invariant. They are $(A \ra A-1,~~ B \ra B+1)$, and $(A \ra A-1,~~ B \ra
-B)$. As explained before in last two examples, the second transformation requires
an extension of the range to ($-\infty,\infty$). The new singular potential
generated for this case is given by
\be
\wti{V}_-(x,A,B) = \[ -A(A-1) \sech^2x +B(B+1) \cosech^2x +(A-B)^2\]
    +4~B \coth x~\frac{x}{|x|}~\delta(x)~.
\ee
The shape invariance condition obeyed by this potential is given by
\be
\wti{V}_+(x,A,B) = \wti{V}_-(x,A+1,-B)+(A-B)^2 - (A-1+B)^2~~,
\ee
and the eigenvalues and eigenfunctions are given by
\br
&& E_0=0~, \nn \\
&& \psi_0 \propto \cosh^{-A}x~\sinh^Bx~,\nn \\
&& E_1=(A-B)^2-(A-1+B)^2~, \nn \\
&& \psi_1 \propto \cosh^{-A}x~\sinh^{-B-1}x
~[(2B-1)\cosh^2x+1]~,\nn \\
&& E_2=(A-B)^2-(A-2-B)^2~,  \nn \\
&& \psi_2 \propto \cosh^A x ~\sinh^B x
~[(4B+2)\cosh^4x-(6B+3)\cosh^2x+1]
~,\nn \\
&& E_3=(A-B)^2-(A-3+B)^2~,  \nn \\
&& \psi_3 \propto
\cosh^Ax ~\sinh^{-B-1}x
~\[(8B^2-16B+6)\cosh^6x-(12B^2-32B+13)\cosh^4x \right.\nn \\
&&\left. \hspace{2.5in} -(20B-8)\cosh^2x-1\] ~.%
\er
Thus, the general formula for eigenvalues is
\be\label{PT-IIenergy}
E_n=(A-B)^2-(A-n-(-1)^n B)^2~~.
\ee
Again, to steer clear of the level crossing problem, we must have $E_n>E_{n-1}$.
This leads to the constraint $-\half < B <\half$; which, as stated earlier, is the
same constraint that one needs for the normalizability of the wavefunction at the
origin and for an effective communication between two halves of the $x$-axis. \sss

%%%%%%%
\noindent{\bf 4. Potential Algebra:}\sms

So far, we have discussed three types of new solvable singular potentials. We will
now derive the potential algebra underlying them. We will show that the algebra
based on the generators $\{J_+,J_-,J_3\}$ is non-linear
\cite{Wu,Alhassid,Gangopadhyaya_proc,ASIM,Balantekin} . Potential algebras provide
an alternative way of getting the eigenvalues by algebraic means.

Consider the following ansatz:
\be
\label{cc}
    \jp =  c^{-1}\,\Ad\(x, \alpha(N), \beta(N)\) ~~,~~
    \jm =  \A\(x, \alpha(N), \beta(N)\) c~~,~~
    \j3 = N \equiv c^{\dagger} c~~,
\ee
where $c, c^{\dagger}$ and $c^{-1} $ are three operators satisfying
$[c,c^{\dagger}]=1$, and $ c\, c^{-1} = c^{-1}\, c = 1$. An example of such
operators is given by $c = e^{i\phi}, ~c^{-1}=e^{-i\phi}$ and $c^{\dagger} = \id \,
e^{-i\phi}$, where $\phi$ is some arbitrary real variable. The operators $\A$ and
$\Ad$ of eq. ( \ref{cc}) are obtained from eq. (\ref{hpm}) via the substitution $a_0
\equiv \{A,B\} \ra \{\alpha(N),\beta(N)\}$, where $\alpha$ and $\beta$ are real,
arbitrary functions to be determined later. We can readily check that
\be
\label{pot-alg}
    \lb \j3,J_{\pm} \rb = \pm J_{\pm} ~,
    ~\lb J_{+}, J_{-} \rb = -R(J_3) \equiv g(\alpha(N), \beta(N) )
    -  g(\alpha(N-1), \beta(N-1) ) ~.
\ee
The last commutation relation is a consequence of the algebraic shape invariance condition
\cite{ASIM}
\be \label{asi}
 H_+(x, \alpha(N),\beta(N)) - H_-\,(x, \alpha(N-1),\beta(N-1))
 = g(\alpha(N-1), \beta(N-1)) - g(\alpha(N), \beta(N))~,
\ee
which is the operatorial ``twin'' of the classical shape invariance condition eq. (\ref{sipv})
obtained via the mappings $\{A_0,B_0\} \ra\{\alpha(N),\beta(N)\}$ and respectively
$\{A_1,B_1\}\ra\{\alpha(N-1),\beta(N-1)\}$.

The functions $\alpha(N)$ and $\beta(N)$ are determined by requiring that the change
$\alpha(N) \ra \alpha(N-1)$ and $\beta(N) \ra \beta(N-1)$ correspond to the change
of parameters $a_0\ra a_1$. For example, $\alpha(N) = A - N$ corresponds to a
translational change of parameters $A_0 \ra A_1 = A_0 +1$, because $\alpha(N-1) =
\alpha(N) +1$. Similarly, $\beta(N) = (-1)^N$ corresponds to the reflection $B_0 \ra
B_1 = -B_0~$, since $~\beta(N-1) = -\beta(N)$.

{}For any shape invariant potential, we know the function $g(\alpha, \beta)$, which
explicitly gives the potential algebra (\ref{pot-alg}).
{}From its representations, we can obtain the energy spectrum for the given problem.

To find a representation of the potential algebra, let us consider a set of
eigenvectors common to both $H_-=J_+ J_-$ and $J_3 =N$ denoted by $\{\v n\>,
n=0,1,\ldots\}$. The action of $J_+,~J_-$ and $J_3$ on this basis is given by
\be
  J_+ \v n \> = a(n+1) \v n +1 \>~,~~ J_- \v n \> = a(n) \v n -1\>~,~~
  J_3 \v n \> = n \v n \>~~.
\ee
Here we have chosen, without any loss of generality, the coefficients $a(n)$ to be real. Note
that since $J_-\v 0 \> = 0$, we have the initial condition $a(0)=0$. There
is a connection between the coefficients $a(n)$ and the eigenspectrum of the
Hamiltonian. Observe that
\be
H_-(x,\alpha(N-1),\beta(N-1))\, \v n \> = J_+ J_- \v n \> = a^2(n) \v n \>~.
\ee
Therefore, in order to find the spectrum of the Hamiltonian we have to determine the
coefficients $a^2(n)$. This can be done by projecting the last equation from
(\ref{pot-alg}) on $\v n \>$ and solving the resulting equation recursively. Thus,
we obtain $a^2(n) - a^2(n+1) = g(n) - g(n-1)$ having the solution $a^2(n) = g(-1)
-g(n-1)$. Here we have denoted $g(n) \equiv g(\alpha(n),\beta(n))$. But $a^2(n)$
corresponds to the eigenvalues of the Hamiltonian $H_-(x,\alpha(N-1),\beta(N-1))$,
or ``classically'' speaking to the shifted set parameters $a_1$. Therefore the
eigenenergies of the initial Hamiltonian $H_-(x,\alpha(N),\beta(N))$ (corresponding
to the set of parameters $a_0$) are
\be \label{en}
E_n = g(\alpha(0),\beta(0)) - g(\alpha(n),\beta(n))~.
\ee
We make contact with eq. (\ref{eq4}) by observing that $\{\alpha(n-k),\beta(n-k)\}\equiv a_k$.
\sss

\noindent{\bf (a) The harmonic oscillator.}

To show how our procedure works, it is instructive to build explicitly the potential algebra of
the harmonic oscillator. The superpartner potentials $V_-$ and $V_+$ are given in eqs. (\ref{vr})
and respectively (\ref{vp}). Under the change of parameters $\{l,\omega\} \ra \{l-1, \omega\}$ we
have the following shape invariance condition
$$
H_+(x,l,\omega) =
H_-(x,l-1,\omega) + (-2 \omega\,(l-1)) - (-2\omega\,l)~.
$$
To build the potential algebra, first we find the functions $\alpha$ and $\beta$ associated with
the above change of parameters. We have immediately $\alpha(N) = l+ N~,~~ \beta(N) = \omega$.
Next, we can build the concrete realization of the potential algebra using the ansatz (\ref{cc})
and the superpotential from (\ref{wx}). The resulting generators
\be
J_+ = c^{-1}\, \( -\frac{d}{dx} + \frac{1}{2}\,x\,\omega + \frac{l+N}{x}\)~,~~
J_- = \( \frac{d}{dx} + \frac{1}{2}\,x\,\omega + \frac{l+N}{x} \)\,c ~,~~
J_3 = N \equiv c^{\dagger} c~,
\ee
satisfy the ``canonical'' commutation relations (\ref{pot-alg}), where the function $g$ is given
by $g(N) \equiv g(\alpha(N),\beta(N))=-2\omega\,(l+N)$. Finally, using the formula (\ref{en}) we
get the spectrum $E_n = g(0)-g(n) = 2\omega\,n$, which is exactly what we have expected.

Next, let us consider the new singular shape invariant potential corresponding to the change of
parameters $ \{l,\omega\} \ra \{-l,\omega\}$. In this case $\alpha(N) = -(-1)^N l$ and $\beta(N)
= \omega$. From eq. (\ref{cc}) we get
\be
J_+ = c^{-1}\, \( -\frac{d}{dx} + \frac{1}{2}\,x\,\omega - \frac{(-1)^N l}{x}\)~,~~
J_- = \( \frac{d}{dx} + \frac{1}{2}\,x\,\omega - \frac{(-1)^N l}{x} \)\,c ~,~~
J_3 = N \equiv c^{\dagger} c~.
\ee
The commutation relations (\ref{pot-alg}) together with the algebraic shape
invariance condition (\ref{asi}) yield in this case $[J_+,J_-]= -\omega\,(-2(-1)^N
l+1)$ from where we get $g(N)=\omega\, ((-1)^N l -N)$. Therefore, the resulting
eigenspectrum (\ref{en}) is $E_n=\omega\,n +\omega\,l\,(1-(-1)^n)$.

\sss

\noindent{\bf (b) The P\"oschl-Teller I potential.}

We build the algebraic model for the new shape invariant  P\"oschl-Teller I like
potential by taking into account that corresponding to the change of parameters
$\{A,B\} \ra \{A +1, -B\}$ we have $\alpha(N) = A - N$ and $\beta(N) = (-1)^N B$.
Then, using the superpotential (\ref{wpt1}) one gets the following expressions for
the generators of the associate potential algebra
\br
J_+ \eq c^{-1}\, \( -\frac{d}{dx} + (A -N ) \tan x - (-1)^N B \cot x \)~,\nn \\
J_- \eq \( -\frac{d}{dx} + (A -N ) \tan x - (-1)^N B \cot x \) \,c~,~~
J_3 = N \equiv c^{\dagger} c~.
\er
Using as before the algebraic shape invariance condition (\ref{asi}) we obtain in this case
$[J_+,J_-]= - (A+N+1+(-1)^{(N+1)} B)^2 + (A+N+(-1)^N B)^2 $. Therefore we get $g(N) = -(A+N
+(-1)^N B)^2$ and the corresponding eigenspectrum $E_n = g(0) - g(n) = -(A+B)^2 + (A-n+(-1)^n B
)$.

\sss

\noindent{\bf (c) The P\"oschl-Teller II potential.}

For the new the P\"oschl-Teller II potential like case, to the change of parameters
$\{A,B\} \ra \{A -1, -B\}$ we have $\alpha(N) = -(A - N)$ and $\beta(N) = (-1)^N B$
and the corresponding algebra is therefore generated by
\br
J_+ \eq c^{-1}\, \( -\frac{d}{dx} + (-A + N ) \tanh x - (-1)^N B \coth x \)~,\nn \\
J_- \eq \( -\frac{d}{dx} + (-A + N ) \tanh x - (-1)^N B \coth x \) \,c~,~~
J_3 = N \equiv c^{\dagger} c~.
\er
In the above representation the explicit form of the superpotential (\ref{wpt2}) was
taken into account. The commutation relations (\ref{pot-alg}) together with the
algebraic shape invariance condition (\ref{asi}) yield in this case $g(N) = (-A+N
-(-1)^N B)^2$. Using (\ref{en}), one obtains as expected, the eigenspectrum for this
potential $E_n = g(0) - g(n) = (A+B)^2 - (-A + n - (-1)^n B )^2$.\sss

\noindent{\bf 5. Conclusions and Comments:}\sms

We have generated several new shape invariant potentials on the whole line starting
from well known potentials on the half line. To ensure continuity and
differentiability of the superpotential, our procedure requires a regularisation at
the origin. This extension not only maintains shape invariance, it also allows the
possibility of a new transformation  among parameters ($B \rightarrow -B$) that was
not allowed on the half-axis. This transformation results in new superpotentials,
albeit singular, that are defined over the entire real axis and have richer spectra
than those defined over half-axis. It is shown further that the eigenspectra of
these new real singular shape invariant potentials may also be derived from a
nonlinear potential algebra.

Since we have obtained and discussed the exact eigenvalues and eigenfunctions of
three new singular potentials using the machinery of supersymmetric quantum
mechanics, it is of interest to ask what one gets in the WKB approximation. Let us
recall that Comtet et al. have shown the exactness of the SWKB quantization
condition\cite{Comtet,dutt}
$$
\int_{x_1}^{x_2} \sqrt{E_n-W^2}~dx = n\pi \hbar~~.
$$
for all known shape invariant problems with unbroken SUSY where parameters are
related by $a_1=a_0+\delta$ \cite{Barclay2}.

For broken SUSY, Inomata and Junker\cite{Inomata} gave the quantization condition
\br
\label{Inomata}
\int_{x_1}^{x_2} \sqrt{E_n-W^2}~dx \eq [n+1/2]\pi \hbar~~.  \nonumber
\er
For both cases, the turning points $x_1, x_2$ are solutions of $W^2(x)=E_n$.

Our new singular potentials allow a change of parameters that, if considered in
half-axes only, leads the system to alternate through unbroken and broken phases of
supersymmetry as $a_{k} \ra a_{k+1}$. It is interesting to note that the spectrum of
these singular potentials can be derived, using somewhat more complex but exact
quantization condition which alternates between the broken and unbroken SUSY cases:
\br
\label{comtet}
  \int_{x_1}^{x_2} \sqrt{E_n-W^2}~dx \eq [n+ {1/2} P_n]~~,  \nonumber
\er
where $P_n$ is equal to $ [1-(-1)^n]/2 $.

%******************************************************
%                ACKNOWLEDGMENT
%******************************************************

Partial financial support from the U.S. Department of Energy is gratefully
acknowledged. R.D. would like to thank the Department of Atomic Energy, Government
of India for a research grant, and the Physics Department of the University of
Illinois at Chicago for warm hospitality.

\sss\sss

\noindent{\bf Appendix A:}~ In this appendix, we show that shape invariance of our
new potentials is maintained during the process of extending the domain to the whole
real axis and introducing the moderating factor $f(x,\epsilon)$. Let us recall that
our old superpotential $W(x,A,B)$ is of the form $A~ \Phi(x) + B~ \[ \Phi(x)
\]^{-1}$, where the function $\Phi(x)$ is $x, \tan x$ or $\tanh x$ for harmonic oscillator,
P\"oschl-Teller I, and P\"{o}schl-Teller II respectively%Footnote begins from here
\footnote{The change of parameters associated  with
shape invariance in these potentials are of the form
\be
A \lra \widetilde{A}=\left\{
    \begin{array}{ll} A  & \\ A+1 & ~~~~{\rm and }~~~~B\lra \widetilde{B}=-B~.  \\ A-1 & \end{array}
\right. .
\ee }
%Footnote ends here

Note that in all cases,
$\[ \Phi(x) \]^{-1} \lra 1/x$ at the origin. $W(x,A,B)$ is replaced by a regularized, continuous superpotential
$\widetilde{W}(x,A,B,\epsilon)$ given by
\begin{equation} \label{Wt1}
\widetilde {W}(x,A,B,\epsilon) = W(x,A,B)~f(x,\epsilon)~,
\end{equation}
where $f(x,\epsilon)$ is unity everywhere except in a small region
of order $\epsilon$ around $x=0$. One such function $f(x,\epsilon)$ is given by
$\tanh^2 \( x/\epsilon \)$. In the limit $\epsilon \ra 0$, we assume that the $f(x,\epsilon) \ra 1$ and
$\frac{d f(x,\epsilon)}{dx} \ra 2  \frac{x}{|x|}  \delta(x)$.
The potentials $\widetilde {V}_\mp(x,A,B,\epsilon)$ corresponding to the superpotential $\widetilde {W}(x,A,B,\epsilon)$ are then given by
\be
\widetilde{V}_{\mp}(x,A,B,\epsilon) = W^2(x,A,B)f^2(x,\epsilon) \mp
\( \frac{d W(x,A,B)}{dx} ~ f(x,\epsilon) + \frac{d f(x,\epsilon)}{dx} ~  W(x,A,B) \) \nonumber ~~.
\ee
Now
\br
\widetilde{V}_+(x,A,B,\epsilon) - \widetilde{V}_-(x,\widetilde{A},\widetilde{B},\epsilon) \eq
V_+(x,A,B) - V_-(x,\widetilde{A},\widetilde{B}) + \frac{d f(x,\epsilon)}{dx} ~
\( W(x,A,B) ~+~ W(x,\widetilde{A},\widetilde{B}) \)
\nonumber \\
\eq R(A,B) + \frac{d f(x,\epsilon)}{dx} ~ \( B + \widetilde{B} \)  \[ \Phi(x) \]^{-1}  \nonumber \\
\eq R(A,B)~~,
\er
where we have used the limits of $f$ and $f'$ and $\widetilde{B}=-B$.
This establishes the shape invariance of the regularized superpotential.
\newpage
%-----------------------------------------------------------------------

\end{document}